\documentclass[prl,twocolumn,showpacs,preprintnumbers,amsmath,amssymb]{revtex4}

\usepackage{mathrsfs}
\usepackage{graphicx}
\usepackage{dcolumn}
\usepackage{bm}

\begin{document}


\title{Fluctuation Theorem with Information Exchange: Role of Correlations in Stochastic Thermodynamics}
\author{Takahiro Sagawa$^{1,2}$}
\author{Masahito Ueda$^{3}$}
\affiliation{$^1$The Hakubi Center for Advanced Research, Kyoto University, Yoshida-Ushinomiya-cho,
Sakyo-ku, Kyoto 606-8302, Japan \\
$^2$Yukawa Institute for Theoretical Physics,  Kyoto University, Kitashirakawa Oiwake-Cho, 606-8502 Kyoto, Japan \\
$^3$Department of Physics, University of Tokyo, 7-3-1 Hongo, Bunkyo-ku, Tokyo 113-0033, Japan
}
\date{\today}

\begin{abstract}
We establish the fluctuation theorem in the presence of information exchange between a nonequilibrium system and other degrees of freedom such as an observer and a feedback controller, where the amount of information exchange is added to the entropy production. 
The resulting generalized second law  sets the fundamental limit of energy dissipation and energy cost during the information exchange.
Our results apply not only to feedback-controlled processes but also to a much broader class of information exchanges, and provides a unified framework of nonequilibrium thermodynamics of  measurement and feedback control.
\end{abstract}


\maketitle

\textit{Introduction.}
Thermodynamics of information processing has attracted great interest ever since  Maxwell's seminal work~\cite{Maxwell,Szilard,Brillouin,Landauer,Bennett,Demon,Maruyama,Maroney0,Sagawa-Ueda5}.
A number of researchers have discussed  various aspects on the relationship between information and thermodynamics~\cite{Lloyd1,Zurek2,Shizume,Landauer2,Goto,Piechocinska,Lutz1,Allahverdyan3,Horhammer,Barkeshli,Norton,Maroney,Turgut,Sagawa-Ueda2,Vedral,Nielsen,Touchette,Zurek1,Touchette2,Cao1,Kim,Sagawa-Ueda1,Cao2,Touchette3,Jacobs,Sagawa-Ueda3,Ponmurugan,Suzuki,Horowitz1,Morikuni,SWKim,Ito,Horowitz2,Abreu,Jarzynski5,Pekola,Horowitz3,Granger,Brandes,Esposito2,Abreu3,Sagawa-Ueda4,Munakata,Esposito,Abreu2,Lahiri,Lutz,Sagawa2,Fetio,Jarzynski6}.  
Such a field of research might be called  ``information thermodynamics.''
Recently, information processing at the level of thermal fluctuations has been  experimentally realized in small thermodynamic systems such as colloidal particles~\cite{Lopez,Toyabe,Berut}.  
Energetic and entropic costs of information processing are vital for designing and controlling nanomachines and nanodevices in thermally fluctuating environments.

A key concept in information thermodynamics is a correlation between two subsystems, which is characterized by the mutual information~\cite{Shannon,Cover-Thomas}.  
If the subsystems are statistically independent, the mutual information vanishes and the entropy is additive, i.e., the Shannon entropy of the total system is given by the sum of those of the subsystems.  
In the thermodynamic limit, a correlation between two subsystems is negligible, and therefore the mutual information vanishes to the leading order.   This is the reason why the entropy is additive in conventional thermodynamics.  In contrast, in small thermodynamic systems, the mutual information can take a positive value and serves as a resource of the work or the free energy through feedback control, as illustrated in Maxwell's gedankenexperiment~\cite{Maxwell}.
The entropy of a system can be decreased without any heat dissipation if we use the correlation as a resource of the entropy decrease, although, in the conventional thermodynamics,  the entropy of the system is decreased only in the presence of heat dissipation. 

Before developing a general theory,  let us consider the Szilard engine~\cite{Szilard} to elucidate the physical meaning of the correlation.
A single-particle gas is enclosed in a box that is in contact with a heat bath at inverse temperature $\beta$.
We consider cyclic processes in which the initial and final states of the gas are in thermal equilibrium with the same volume of the box.  
If we do not know the position of the particle, we cannot extract a positive amount of work from the gas because of the second law of thermodynamics (i.e., the Kelvin principle).  However, if we have one bit  ($= \ln 2$ nat in the natural logarithm) of information about the initial position of the particle, we can extract $\beta^{-1} \ln 2$ of work by means of feedback control.
To obtain the information, we insert a barrier at the center of the box, and measure whether the particle is in the left or right side (see also Fig.~1).  The measurement outcome is recorded in a memory device. 
Here, the ``information'' means the correlation between the position of the gas and the memory, which is described by $\ln 2$ of the mutual information.  It is then used as a resource of the work through an isothermal expansion of the left or the right box.
\begin{figure}[htbp]
 \begin{center}
 \includegraphics[width=70mm]{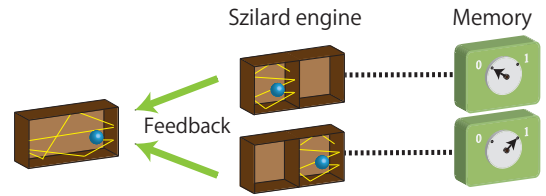}
 \end{center}
 \caption{(color online). The Szilard engine and  memory which are initially correlated with one bit ($= \ln 2$ nat) of the mutual information. We can then use this information to extract $\beta^{-1} \ln 2$ of work by means of feedback control. Here the information is used for deciding in which direction we perform an isothermal expansion.} 
\end{figure}



The Szilard engine  is restricted to  the special setup with two boxes.  How should fundamental laws of thermodynamics  be modified in the presence of a more general correlation between the system and other degrees of freedom?  This question involves a broad class of information processes of thermodynamic systems including the situations of measurement and feedback control.

In this Letter, we  generalize the fluctuation theorem (FT)~\cite{Jarzynski1,Crooks1,Crooks2,Jarzynski2,Seifert,Kawai,JarzynskiR,SeifertR} and the second law of thermodynamics (SL) by explicitly taking correlations into account, where the entropy production (EP) and the mutual information are treated on an equal footing.
Our setup includes measurement and feedback control as important special cases, where they are treated in the same framework. 
As a corollary, we obtain a generalized FT that applies to measurement processes.
In addition, our previous results, such as a generalized FT for feedback control~\cite{Sagawa-Ueda3} and the minimal energy cost for measurement~\cite{Sagawa-Ueda2}, are reproduced as special cases of our results.
Our results are valid not only in Langevin systems but also far from equilibrium situations of classical stochastic dynamics, because they can be derived on the basis of the detailed FT~\cite{Crooks1,Crooks2,Jarzynski2,Seifert}.

\textit{Setup.}
Suppose that a classical stochastic system $\bm X$ is in contact with multi-heat baths labeled by $k = 1, 2, \cdots$ at inverse temperatures  $\beta_k$.
System $\bm X$ may be driven to far from equilibrium by changing external parameters. 
We can extract  work from $\bm X$ through such external parameters. 
We assume that the time evolution of $\bm X$ is described by a classical stochastic dynamics from $t=0$ to $t = \tau$ along trajectory $X_F$.
Let $x$  ($x'$) be the initial (final) phase-space point of $\bm X$, and $P^i_F[x]$ ($P^f_F[x']$) be the corresponding probability distribution.
The EP in $\bm X$ and the baths is then given by
\begin{equation}
\sigma := \Delta s - \sum_k \beta_k Q_k,
\label{entropy_production}
\end{equation}
where $\Delta s := (- \ln P^f_F[x'] ) -  (- \ln P^i_F[x])$  is a change in the stochastic entropy and $Q_k$ is the heat absorbed by the system from the $k$th bath.
We note that $-\beta_k Q_k$ is regarded as the entropy change in the $k$th bath.

In addition to the heat baths, system $\bm X$ interacts with another system $\bm Y$.
We assume that $\bm Y$ does not evolve in time during the interaction.  
This assumption applies not only to the standard setup of Maxwell's demon but also to a broader class of information processing. 
System $\bm X$ then evolves depending on the state $y$ of $\bm Y$  (see also Fig.~2).
The conditional probability of  $X_F$ being realized under the initial condition of  $x$ depends on $y$ such that the conditional probability is given by $P_F [X_F | x, y]$.

\begin{figure}[htbp]
 \begin{center}
 \includegraphics[width=35mm]{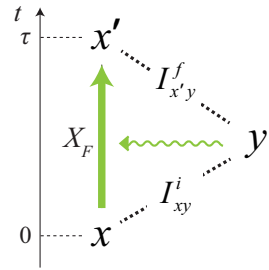}
 \end{center}
 \caption{(color online). Interaction between $\bm X$ and $\bm Y$.  System $\bm X$ evolves from $x$ to $x'$ in such a manner that depends on the information about $y$, while $y$ does not evolve in time. } 
\end{figure}

We also assume that  $x$ may be correlated with $y$ in the initial and final states.   
Let  $P^i_F[x,y]$ ($P^f_F[x',y]$) be the initial (final) joint probability distribution of $x$ ($x'$) and $y$, where  $P^i_F[x] = \int dy P^i_F[x,y]$ ($P^f_F[x'] = \int dy P^f_F[x',y]$) and  $P_F[y] = \int dx P^i_F[x,y] = \int dx' P^f_F[x',y]$ are the corresponding marginal distributions.
 If $P^i_F[x,y] = P^i_F[x]P_F[y]$, $x$ and $y$ are statistically independent.  Otherwise, they are correlated.
We characterize the initial and final correlations between the two systems by
\begin{equation}
I^i_{xy} := \ln \frac{P^i_F[x,y]}{P^i_F[x]P_F[y]}, \ I^f_{x'y} := \ln \frac{P^f_F[x',y]}{P^f_F[x']P_F[y]}.
\label{mutual}
\end{equation}
If $x$ and $y$ ($x'$ and $y$) are not correlated, $I^i_{xy}$ ($I^f_{x'y}$) vanishes.  The ensemble averages of~(\ref{mutual}) ($\langle I^i_{xy} \rangle$ and $\langle I^f_{x'y} \rangle$) give the mutual information~\cite{Shannon,Cover-Thomas}.  We also refer to $I^i_{xy}$ and $I^f_{x'y}$ as the mutual information. 
Then, $\Delta I := I^f_{x'y} - I^i_{xy}$ gives the change in the mutual information during the dynamics.

As discussed in detail later, this setup includes both cases of  measurement and feedback control.
In the case of measurement, $\langle I^i_{xy} \rangle = 0$  and $\langle I^f_{x'y} \rangle >0$ hold, where   $\langle I^f_{x'y} \rangle$ describes the obtained information.  In contrast, in the case of feedback control, $\langle I^i_{xy} \rangle>0$, which is the resource of the work  and the free energy, and $\langle I^f_{x'y} \rangle (< \langle I^i_{xy} \rangle)$ is the remaining correlation after the feedback control.

\textit{Main results.}
In the absence of the initial or final correlations, EP~(\ref{entropy_production}) satisfies the integral FT (or the Jarzynski equality)   $\langle e^{-\sigma} \rangle = 1$~\cite{Jarzynski1,Crooks2,Seifert}, where $\langle \cdots \rangle$ describes the ensemble average over all microscopic trajectories.  
In contrast, in the presence of information processing with initial and final correlations, the integral FT  is generalized as
\begin{equation}
\langle e^{-\sigma + \Delta I} \rangle = 1,
\label{Jarzynski2}
\end{equation}
where we assume that
\begin{equation}
P^i_F[x,y] \neq 0 \ \ {\rm for \ any} \ (x,y).
\label{assumption}
\end{equation}  
Equality~(\ref{Jarzynski2}) is the main result in this Letter, which will be proved later.
By using the convexity of the exponential function $e^{\langle x \rangle} \leq \langle e^x \rangle$, Eq.~(\ref{Jarzynski2}) leads to 
\begin{equation}
\langle \sigma \rangle \geq  \langle \Delta I \rangle.
\label{second2}
\end{equation}
Equality~(\ref{Jarzynski2}) and inequality~(\ref{second2}) imply that we can control EP $\sigma$ in the subsystem by changing the correlation.
In the absence of initial or final correlations, Eq.~(\ref{Jarzynski2}) and inequality~(\ref{second2}) reduce to the conventional FT and SL, respectively.
Inequality~(\ref{second2}) holds without assumption~(\ref{assumption}), while Eq.~(\ref{Jarzynski2}) does not, as shown later.

In the presence of a single heat bath at inverse temperature $\beta$, inequality~(\ref{second2}) implies the minimal energy dissipation
\begin{equation} 
- \beta \langle Q \rangle \geq - \langle \Delta s \rangle + \langle \Delta I \rangle,
\label{heat_bound}
\end{equation}
where $-\langle Q \rangle$ is the heat transfered from $\bm X$ to the bath. 
If $\langle \Delta I \rangle = 0$, inequality~(\ref{heat_bound}) reduces to $-  \beta \langle Q \rangle \geq - \langle \Delta s \rangle$,
which leads to the celebrated Landauer principle and its generalizations~\cite{Landauer,Shizume,Piechocinska,Lutz1,Allahverdyan3,Horhammer,Barkeshli,Norton,Maroney,Turgut,Sagawa-Ueda2}.
We note that $\langle \Delta s \rangle$ is a change in the total Shannon entropy.

We next consider the energy cost for the information exchange with a single heat bath.
Let $E[x; t]$ be the Hamiltonian of $\bm X$ and $E_{\rm int}[x,y; t]$ be the interaction Hamiltonian between $\bm X$ and $\bm Y$.
We note that $E$ and $E_{\rm int}$ can be dependent explicitly on time $t$ due to a change in external parameters.
The first law of thermodynamics is given by
\begin{equation}
\Delta E + \Delta E_{\rm int} = W + Q,
\label{first_law}
\end{equation}
where $\Delta E := E[x'; \tau] - E[x; 0]$, $\Delta E_{\rm int} := E_{\rm int}[x', y; \tau] - E_{\rm int}[x,y; 0]$, and $W$ is the work performed on the system.   
From Eqs.~(\ref{heat_bound}) and (\ref{first_law}), we obtain
\begin{equation} 
\langle W \rangle \geq \Delta F_{\rm eff}   + \langle \Delta E_{\rm int} \rangle + \beta^{-1} \langle \Delta I \rangle, 
\label{work_bound}
\end{equation}
where $F_{\rm eff} (t) := \langle E (t) \rangle - \beta^{-1} \langle s (t) \rangle$ is an effective free energy of $\bm X$~\cite{Esposito2}, and $\Delta F_{\rm eff} := F_{\rm eff} (\tau) - F_{\rm eff} (0)$.
We note that there is no thermodynamic restriction on the value of $\langle \Delta E_{\rm int} \rangle$ as is the case for  $\langle \Delta E \rangle$.
Inequality~(\ref{work_bound}) shows that the correlation-induced energy cost $\beta^{-1} \langle \Delta I \rangle$ is added to the conventional bound in thermodynamics.

A crucial point of our setup is that the entropy of $\bm X$ can be decreased without any heat flow.  In conventional thermodynamics, $\langle \Delta s \rangle \geq \beta \langle Q \rangle$, where the negative heat flow allows the entropy to decrease.  In contrast, in our setup, $\langle \Delta s \rangle \geq \langle \Delta I \rangle$ if  $\langle Q \rangle = 0$, where the negative mutual-information change is the resource of the entropy decrease in $\bm X$.
Such an information-energy balance is based on the dynamics characterized by Fig.~2, where one of the two systems does not evolve in time.  The generalization of our results to more involved situations, where the two systems can influence each other and evolve in time, is an interesting future challenge.

We now discuss important special cases: measurement and feedback control.

\textit{Measurement.}
Let $\bm X$ be a measuring system (a demon) and $\bm Y$ a measured system  (see Fig.~3 (a) for a special case).  We assume that $\bm X$ is initially not correlated with $\bm Y$.  Then, $\bm X$ performs a measurement on the value of $y$.  In the final state, $I := I^f_{x'y}$ characterizes the information gain by  $\bm X$ that is positive because of  $\langle \Delta I \rangle = \langle I \rangle> 0$.
In this case, Eq.~(\ref{Jarzynski2}) reduces to the generalized integral FT for the measurement process:
\begin{equation}
\langle e^{-\sigma + I} \rangle = 1.
\end{equation}
 Consequently, inequality~(\ref{second2})  reduces to  $\langle \sigma \rangle \geq \langle I \rangle$,
which means that an additional EP is accompanied by the measurement. 
 Our result is consistent with Bennett's observation that the energy dissipation can be zero during the measurement process with a single heat bath~\cite{Bennett}.  In fact, if both  $\langle \Delta s \rangle$ and $\langle I \rangle$ equal the Shannon information obtained by the measurement, inequality~(\ref{heat_bound}) reduces to $- \langle Q \rangle \geq 0$, which is the case that Bennett considered.  In general, the minimal heat dissipation is given by $-\beta  \langle Q \rangle \geq - \langle  \Delta s \rangle + \langle I \rangle$.
The work needed for the measurement is bounded as $\langle W \rangle \geq \Delta F_{\rm eff}   + \langle \Delta E_{\rm int} \rangle + \beta^{-1} \langle I \rangle$, where $\beta^{-1} \langle I \rangle$ describes an additional energy cost. A special case of this inequality was obtained in Ref.~\cite{Sagawa-Ueda2}.

\begin{figure}[htbp]
 \begin{center}
 \includegraphics[width=80mm]{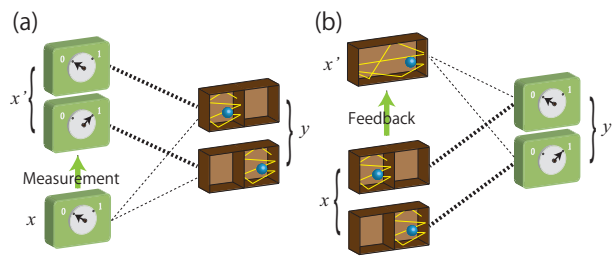}
 \end{center}
 \caption{(color online).  (a) Position measurement on the Szilard engine ($\bm Y$) by a memory system ($\bm X$), where $x$ describes the phase-space point of the memory, and $y$ takes on $0$ or $1$ corresponding to ``left'' or ``right.''  The initial correlation is $\langle I^i_{xy} \rangle = 0$ and the final correlation is $\langle I^f_{x'y} \rangle = \ln 2$.   (b)  Feedback control on the Szilard engine ($\bm X$) by a demon ($\bm Y$), where $x$ describes the position of the particle, and $y$ ($= 0$ or $1$) is the measurement outcome stored in the memory. The initial correlation is $\langle I^i_{xy} \rangle = \ln 2$ and the final correlation is $\langle I^f_{x'y} \rangle = 0$.} 
\end{figure}

\textit{Feedback control.} Let $\bm X$ be a controlled system and $\bm Y$ a controller (a demon)  (see Fig.~3 (b) for a special case). We assume that $\bm X$ is initially correlated with $\bm Y$ with mutual information $I := I^i_{xy}$. By using this initial correlation, $\bm Y$ performs feedback control on $\bm X$ in such a manner that the evolution of $\bm X$ depends on $y$.  The correlation remaining after the feedback is given by $I_{\rm rem} := I^f_{x'y}$.  In this case, Eq.~(\ref{Jarzynski2}) is equivalent to 
\begin{equation}
\langle e^{-\sigma - (I - I_{\rm rem})} \rangle = 1,
\label{Jarzynski_feedback}
\end{equation} 
which is the generalized integral FT for the feedback process.
We note that  a similar equality was obtained in Ref.~\cite{Sagawa-Ueda3}, which is equivalent to the present result with $I_{\rm rem}=0$.
Corresponding to Eq.~(\ref{Jarzynski_feedback}), inequality~(\ref{second2}) reduces to $\langle \sigma \rangle \geq - \langle I  - I_{\rm rem} \rangle$, where $\langle I - I_{\rm  rem} \rangle  > 0$ sets an upper bound of the correlation that is utilized by the demon.
Correspondingly, the work that is extracted from a single heat bath by the demon, denoted as $W_{\rm ext} :=  -W$, is bounded as $\langle W_{\rm ext} \rangle \leq - \Delta F_{\rm eff}   - \langle \Delta E_{\rm int} \rangle + \beta^{-1} \langle  I - I_{\rm rem} \rangle$, where $ \beta^{-1} \langle  I - I_{\rm rem} \rangle$ describes the extractable work on top of the conventional bound.

\textit{Derivation of the main result.}
We now derive Eq.~(\ref{Jarzynski2}) and several related relations.
We first note that $P_F[X_F, y] = P_F [X_F | x, y] P^i_F[x, y]$, where $P_F[X_F , y] $ is the joint probability distribution of realizing trajectory $X_F$ and $y$. 
We then introduce the backward processes, in which the control protocol of external parameters is time-reversed.
For simplicity, we assume that the phase space of the system does not include momentum terms; the generalization of our arguments to situations with momentum terms is straightforward.
In considering the time-reversed trajectory,  the initial probability distribution of the backward process is taken to be the final distribution of the forward process. 
Let $X_B$ be the time-reversed trajectory of $X_F$.   The joint probability distribution of $(X_B, y)$ in the backward processes, denoted as $P_B[X_B, y] $, is given by $P_B[X_B, y] = P_B[X_B | x', y]  P^f_F[x',y]$,
where $P^f_F[x',y]$ is the initial probability distribution of the backward processes and $P_B[X_B | x', y] $ is the conditional probability of realizing $X_B$ under the initial condition of $(x',y)$.

The detailed FT in our setup is given by
\begin{equation}
\frac{P_B[X_B | x', y]}{P_F[X_F | x, y]} = e^{\sum_k \beta_k Q_k},
\label{DFT}
\end{equation}
where we used the assumption that $y$ does not evolve in time.  We note that Eq.~(\ref{DFT}) holds even when $\langle \Delta E_{\rm int} \rangle \neq 0$.
We also note that detailed FT can be proved in the presence of multi-heat baths in several setups.  For example, it has been proved under the assumptions that the total system including the heat baths obeys the Hamiltonian dynamics and that the initial probability distributions of the baths are the canonical distributions~\cite{Jarzynski2}. 
By noting that
\begin{equation}
\begin{split}
&\frac{P_B[X_B, y]}{P_F[X_F, y]} = \frac{P_B[X_B | x', y]}{P_F[X_F | x, y]} \! \cdot \! \frac{P^f_F[x']}{P^i_F[x]}  \! \cdot  \! \frac{P^f_F[x', y] / P^f_F[x']}{P^i_F[x,y] / P^i_F[x]} \\
&=  \frac{P_B[X_B | x', y]}{P_F[X_F | x, y]} \cdot \frac{P^f_F[x']}{P^i_F[x]} \cdot \frac{P^f_F[x',y]}{P^f_F[x']P_F[y]} \cdot \frac{P^i_F[x]P_F[y]}{P^i_F[x,y]},
\end{split}
\label{eq}
\end{equation}
we obtain
\begin{equation}
\frac{P_B[X_B, y]}{P_F[X_F, y]} = e^{-\sigma + \Delta I},
\label{FT2}
\end{equation}
which is the detailed FT in the presence of information processing.
We note that $\sigma - \Delta I$ can be regarded as the total EP in the composite system $\bm{XY}$ and the baths, and therefore the total system satisfies the conventional FT.
Equality~(\ref{FT2}) implies the trade-off relation between the mutual-information change and EP in the subsystem.

By using $dX_F = dX_B$, $\int dX_B dy P_B[X_B, y] = 1$, and assumption~(\ref{assumption}), we  obtain 
\begin{equation}
\langle e^{-\sigma + \Delta I} \rangle = \int \frac{P_B[X_B, y]}{P_F[X_F, y]} P_F[X_F, y] dX_F dy = 1,
\end{equation}
which implies Eq.~(\ref{Jarzynski2}).
Moreover, we obtain from Eq.~(\ref{FT2}) that
\begin{equation}
\langle \sigma - \Delta I \rangle  = \int P_F[X_F, y] \ln \frac{P_F[X_F, y]}{P_B[X_B, y]}  dX_F dy,
\label{KPB}
\end{equation}
where the right-hand side is the relative entropy between the forward and backward probabilities.
We note that Eq.~(\ref{KPB}) is a generalization of the result in Ref.~\cite{Kawai}.
Inequality~(\ref{second2}) is also confirmed from Eq.~(\ref{KPB}) because of the positivity of the relative entropy.
The equality in (\ref{second2}) is achieved if $P_F[X_F, y] = P_B[X_B, y]$ holds for any $X_F$ and $y$.

We discuss the case in which assumption (\ref{assumption}) is not satisfied.  Let $\mathcal S$ be a set of $(x,y)$ such that $P^i_F[x,y] \neq 0$ holds for $(x, y) \in \mathcal S$.  We do not observe any event outside $\mathcal S$, because it has the zero probability.  We then obtain
\begin{equation}
\begin{split}
\langle e^{-\sigma + \Delta I} \rangle & := \int_{(x,y) \in \mathcal S} \frac{P_B[X_B, y]}{P_F[X_F, y]} P_F[X_F, y] dX_F dy  \\
&= \int_{(x,y) \in \mathcal S} P_B[X_B, y] dX_B dy,
\end{split}
\end{equation}
which is not necessarily unity.   In other words, the left-hand side of Eq.~(\ref{Jarzynski2}) does not converge to unity in the limit of $P^i_F[x,y] \to + 0$ for $(x, y) \in \mathcal S$.  In contrast, the right-hand side of Eq.~(\ref{KPB}) converges in the same limit, and therefore Eq.~(\ref{KPB}) and inequality~(\ref{second2}) still hold without assumption~(\ref{assumption}).

\textit{Concluding remarks.}
We have addressed the question of correlations in thermodynamics by deriving  the generalized  integral FT~(\ref{Jarzynski2}) and SL (\ref{second2}).  The generalized SL leads to the minimal heat dissipation~(\ref{heat_bound}) and the minimal energy cost~(\ref{work_bound}) for information exchanges.
As corollaries, we have derived the generalized FT and SL for measurement and feedback-controlled processes in a single framework.
Our results serve as guiding principles for designs of artificial nanomachines and nanodevices; for example, we can judge how efficient nanomachines with information processing can be, by comparing their entropy productions with the lower bounds of inequality~(\ref{second2}).
It is interesting to investigate a fluctuation theorem that only involves the variable of one of the systems that exchange information.  Such a fluctuation theorem has been found only for feedback-controlled processes~\cite{Sagawa-Ueda4}.
Moreover, experimental verifications of our results in small thermodynamic systems merits further study.

\begin{acknowledgments}
This work is  supported by the Global COE Program ``The Next Generation of Physics, Spun from Universality and Emergence'' and ``The Physical Science Frontier'' from MEXT of Japan,  and by the  Grant-in-Aid for Scientific Research on Innovative Areas ``Topological Quantum Phenomena'' (No. 22103005), KAKENHI 22340114, and KAKENHI 21540384.
\end{acknowledgments}

\end{document}